\documentclass[aps,prd,preprint,onecolumn,showpacs]{revtex4-1}
\usepackage{amsmath}
\usepackage{amsthm}
\usepackage{bm}          
\usepackage{dcolumn}     
\usepackage{graphicx}    
\usepackage{multirow}    
\usepackage{rotating}    
\usepackage{supertabular}
\usepackage{afterpage}   
\usepackage{float}       
\usepackage{mathrsfs}    
\usepackage{amsfonts}    
\usepackage{upgreek}     
\usepackage{color}
\usepackage{epstopdf}    
\usepackage{orcidlink}

\hypersetup{
	colorlinks=true,
	citecolor=blue,
	urlcolor=blue,
	linkcolor=blue
}

\begin{document}
	\title{Neutrino Constraints on memory-burdened Primordial Black Holes from Dwarf Spheroidal Galaxies}
	
	\author{Xiu-Hui Tan\orcidlink{0000-0003-2940-6664}\thanks{Corresponding author}}
	\email{tanxh@itp.ac.cn}
	\affiliation{Institute of Theoretical Physics, Chinese Academy of Sciences, Beijing 100190, China}
	\author{Jun-qing Xia}
	\email{xiajq@bnu.edu.cn}
	\affiliation{Department of Astronomy, Beijing Normal University, Beijing 100875, China}
	\affiliation{Institute for Frontiers in Astronomy and Astrophysics, Beijing Normal University, Beijing 100875, China}
	\author{Yu-Feng Zhou}
	\email{yfzhou@itp.ac.cn}
	\affiliation{Institute of Theoretical Physics, Chinese Academy of Sciences, Beijing 100190, China}
	\affiliation{School of Physical Sciences, University of Chinese Academy of Sciences, Beijing 100049, China}
	\affiliation{School of Fundamental Physics and Mathematical Sciences, Hangzhou Institute for Advanced Study, UCAS, Hangzhou 310024, China}
	\affiliation{International Centre for Theoretical Physics Asia-Pacific, Beijing/Hangzhou, China}

\begin{abstract} 
	Dwarf spheroidal galaxies (dSphs) represent prime targets for indirect dark matter (DM) searches due to their substantial DM content and low astrophysical backgrounds. In this work, we conduct a comprehensive search for neutrino signals originating from memory-burdened primordial black holes (PBHs) as DM candidates, utilizing 10 years of publicly available muon-track data from the IceCube Neutrino Observatory. We systematically compile $\mathcal{D}$-factor measurements from four independent literature sources, resulting in a robust sample of 14 dSphs with well-characterized DM distributions. For each dSph, we perform an unbinned maximum-likelihood analysis to evaluate the significance of potential PBH neutrino emission, incorporating $\mathcal{D}$-factor uncertainties through a profile likelihood framework. No significant excess over the background-only hypothesis is detected. Building upon these null results, we derive upper limits on the PBH abundance fraction, assuming a monochromatic mass distribution. Our analysis demonstrates that constraints obtained in the $k=1$ case exhibit significant improvement compared to previous studies, while the $k=2$ case yields less restrictive limits due to limited sensitivity of IceCube in the relevant energy range. Furthermore, our combined analysis of 14 dSphs achieves substantially enhanced sensitivity, as the independent error treatment effectively reduces statistical uncertainties. 
\end{abstract} 
\maketitle 
\section{Introduction}\label{sec:intro} 
For several decades, intriguing objects known as Primordial Black Holes (PBHs), which form from the direct collapse of overdensities in the early Universe, have been proposed as compelling cosmological phenomena \cite{Hawking:1971ei,Carr:1974nx}. Recently, they have attracted renewed attention following the detection of black hole mergers via gravitational waves, as PBHs serve as compelling dark matter (DM) candidates and potential sources for gravitational wave events \cite{Carr:2020xqk, Green:2020jor}. The typical mass range of PBHs spanning from the Planck mass $M_P$ to stellar masses, however, not all PBHs can survive to the present day. On one hand, very heavy PBHs would significantly influence the outcomes of big bang nucleosynthesis (BBN) and the cosmic microwave background (CMB) due to their substantial gravitational effects. On the other hand, once formed, PBHs undergo the semi-classical process known as Hawking Evaporation (HE), which causes them to gradually lose mass. Consequently, because of HE, only PBHs with masses exceeding $\sim 5\times 10^{14}$ g can survive to the present epoch. 

Since asteroid-mass PBHs exhibit many properties characteristic of DM, numerous studies \cite{Bambi:2008kx, Carr:2009jm, Carr:2016hva, DeRocco:2019fjq, Laha:2019ssq, Laha:2020ivk, Carr:2020gox, Chen:2021ngo, Iguaz:2021irx, Tan:2022lbm, Tan:2024nbx, Huang:2024xap} have searched for them via their electromagnetic and cosmic-ray signatures. Recent theoretical developments have introduced the concept of \textit{memory burden} (MB) effect, which emerges from the black hole quantum information paradox and modifies the evaporation process \cite{Dvali:2018xpy, Dvali:2020wft, Dvali:2024hsb}. This effect fundamentally alters the late-stage evolution of PBHs: when a PBH has lost approximately half of its initial mass, it transitions into a distinct phase characterized by significantly suppressed evaporation rates. The underlying mechanism involves a back-reaction from the stored information with their entropy, effectively slowing down the emission process. When the energy of the emitted quanta becomes comparable to the black hole's total energy, the MB effect dominates the evolution of the PBHs. Thus, it prevents light PBHs from undergoing catastrophic evaporation during epochs critical for cosmological evolution. Although the MB effect suppresses the emission of radiation, the viable mass window for PBHs as dark matter candidates has been substantially extended. It is derived that masses as low as $\sim 10^4{~\rm g}$ remain consistent with both BBN constraints and CMB observations, and these light PBHs may still produce observable high-energy emissions. 

These memory-burdened PBHs serve as potential sources of high-energy particle emission, spanning the electromagnetic spectrum and including gamma-rays, neutrinos, and cosmic rays. A growing number of papers discuss this theme \cite{Su:2024hrp, Alexandre:2024nuo, Thoss:2024hsr, Haque:2024eyh, Chianese:2025wrk, Liu:2025vpz, Ambrosone:2026djo}. Neutrinos are very promising probes for exploring the high-energy Universe due to their weak interaction cross-sections with matter, which enables them to propagate cosmological distances without significant attenuation or deflection by magnetic fields. This characteristic allows neutrinos to preserve information about their production mechanisms and source locations, making them invaluable tracers for identifying and studying distant astrophysical accelerators, including potential PBH evaporation sites. 
Many previous studies have investigated neutrino fluxes from PBHs \cite{Halzen:1995hu, Lunardini:2019zob, Dasgupta:2019cae, Chianese:2024rsn, Chaudhuri:2025rcs}. Motivated by the MB effect, utilizing neutrino data provides a promising avenue to probe the properties of memory-burdened PBHs as viable DM candidates. Consequently, this work aims to explore this scenario.

The IceCube Neutrino Observatory is an in-ice detector and detects high-energy neutrinos by capturing the Cherenkov light emitted by relativistic charged secondary particles produced in neutrino-nucleon interactions, traveling through the deep, ultra-clean glacial ice. It also has sensitivity to high-energy atmospheric muons above $\sim$300 TeV. Neutrino interactions observed in the IceCube array generally exhibit either track-like or cascade-like topology. The track-like signal events originate primarily from charged-current interactions of muon (anti-)neutrinos with nucleons, producing energetic muons. From April 2008 to July 2018, IceCube has collected 10 years of muon-track data \cite{Braun:2008bg, IceCube:2021xar}, which has been commonly used in previous research \cite{Sandick:2009bi, IceCube:2023ies, Guo:2023axz, Lu:2024jbq}. In those researches, no significant excess beyond background expectations has been found in DM searches using this dataset. 

If PBHs constitute all or a fraction of the DM, they are expected to cluster within galactic DM halos. Dwarf spheroidal galaxies (dSphs), being the most DM-dominated systems known with exceptionally low astrophysical backgrounds, naturally emerge as prime targets for such a search. In this work, we investigate the scenario where memory-burdened PBHs constitute the DM in dSphs and emit high-energy neutrinos via Hawking radiation. The detection of such neutrinos from dSphs would provide a important signature for PBH DM, as these quiet environments are devoid of conventional astrophysical accelerators capable of generating comparable high-energy signals.

The paper is organized as follows: in Section \ref{sec:model}, we derive the differential neutrino flux from the memory-burdened PBHs. Section \ref{sec:ana} describes the IceCube data and likelihood analysis, while Section \ref{sec:res} presents our results. Finally, our main conclusions are summarized in Section \ref{sec:concl}. Throughout this work, we adopt natural units where $\hbar = c=k_B=1$ and define the reduced Planck mass as $M_P=(8\pi G)^{-1/2}$. 

\section{Model}\label{sec:model} 
\subsection{Memory-burdened PBHs} 
Within a semi-classical framework, Hawking calculated the spectrum of particles emitted by black holes following their gravitational collapse \cite{Hawking:1974rv, Hawking:1975vcx}. The emission rate of PBHs is determined by their mass $M$, angular momentum $J$, and charge $Q$. However, towards the end of their lifetimes, PBHs can be effectively described as Schwarzschild black holes, since the mass typically remains more stable than both charge and angular momentum. The rate of mass loss for a PBH can be expressed as 
\begin{equation} 
	\frac{{\rm d}M_{\rm BH}}{{\rm d}t}=-\epsilon(M_{\rm BH})\frac{M_P^4}{M_{\rm BH}^2}, \label{eq:massloss_sc} 
\end{equation} 
where $\epsilon(M_{\rm BH})$ is the evaporation function that encodes the contribution from all particle degrees of freedom (dof) that a PBH of mass $M_{\rm BH}$ can emit, which can be expressed as 
\begin{equation} 
	\epsilon(M_{\rm BH})=\sum_i \frac{g_i}{128 \pi^3} \int_0^\infty \frac{x\Gamma_{s_i}(x)}{e^x-(-1)^{2s_i}} {\rm d}x. 
	\label{eq:epsilon}
\end{equation} 
$\Gamma_{s_i}$ represents the greybody factor for the particles with spin $s_i$ follows a Fermi-Dirac (or Bose-Einstein) case, which characterizes the backscattering of particles by the gravitational and centrifugal potentials near the black hole horizon, i.e., the absorption probability. $g_i$ denotes the internal dof of a particle species $i$. The dimensionless parameter $x$ is defined as the ratio of particle energy to black hole temperature, $x=E/T$, where $T$ is the Hawking temperature, which is related to the PBH mass as: 
\begin{equation} 
	T=\frac{1}{8 \pi G M} \simeq 10^9 ~{\rm GeV}~\left(\frac{10^{14}{\rm g}}{M_{\rm BH}} \right). 
\end{equation} 
The above equation indicates that a PBH can emit standard model particles with masses $m\lesssim 10^9~{\rm GeV}$ and even degrees of freedom beyond the Standard Model. By integrating Eq. \ref{eq:massloss_sc}, we can exactly solve for the PBH mass as a function of time: 
\begin{equation} 
	M(t) = M_{\rm in}\left(1-\frac{t}{\tau_{\rm SC}}\right)^{1/3}, \label{eq:lossrate_SC} 
\end{equation} 
where $M_{\rm in}$ is the initial mass of the PBH when it forms, and $\tau_{\rm SC}$ denotes the semi-classical lifetime. We can obtain $\tau_{\rm SC}$ by substituting $M_{\rm BH}(\tau_{\rm SC})=0$ in Eq. \ref{eq:lossrate_SC}, 
\begin{equation} 
	\tau_{\rm SC}=\frac{M_{\rm in}^3}{3\epsilon M_p^4}\simeq 2.4\times 10^{-28}\left(\frac{M_{\rm in}}{1{~\rm g}}\right)^3 {~\rm s}. 
\end{equation} 
From this expression, we recover that PBHs with $M_{\rm in}\lesssim 10^9 {~\rm g}$ have evaporated before BBN, whereas PBHs with $M_{\rm in}\gtrsim 10^{15}{~\rm g}$ still survive today. The instantaneous emission rate of a particle $i$ from PBH evaporation, according to Eq. \ref{eq:epsilon}, is given by 
\begin{equation} 
	\frac{{\rm d}^2 N_i}{{\rm d}E{\rm d}t}=\frac{g_i}{2 \pi}\frac{\Gamma_{s_i}(x)}{e^{x}-(-1)^{2 s_i}}. 
	\label{eq:emission_rate}
\end{equation} 

From the semi-classical regime transfer to the MB effect, there are several different scenarios to consider. The evaporation process can proceed in fast, slow, or merger modes in recent researches \cite{Dondarini:2025ktz, Montefalcone:2025akm, Dvali:2025ktz}. The simplest and most straightforward is the fast mode, which assumes that the semi-classical regime ends when 
\begin{equation} 
	M_{\rm BH}=q M_{\rm in}, 
\end{equation} 
where $q$ is a free parameter that determines when the MB effect begins. Since the PBH mass diminishes with time, we have $0<q<1$. Typically, $q=1/2$ is a reasonable choice \cite{Dvali:2018xoc}. The exact numerical choice of $q$ does not affect the qualitative behavior of the evaporation process, but instead determines how the derived bounds map onto the PBH mass at formation. In this work, we focus on the neutrino signals of PBHs from dSphs, where our bounds are largely insensitive to the detailed early-time behavior of the evaporation process. The limits presented here can therefore be straightforwardly reinterpreted within alternative frameworks featuring non-instantaneous suppression, provided the corresponding modification to the late-time evaporation rate is specified. A systematic treatment of such extended scenarios is left for future investigation. 

The information stored in the PBHs back-reacts and extends their lifetime via the entropy $S(M_{\rm PBH})=4\pi GM^2_{\rm PBH}$, causing the decay rate to slow down as 
\begin{equation} 
	\frac{{\rm d}M_{\rm PBH}^{\rm MB}}{{\rm d}t}=\frac{1}{S(M_{\rm PBH})^k} \frac{{\rm d}M_{\rm PBH}}{{\rm d}t}, {\rm with}~k>0, \label{eq:massloss_mb} 
\end{equation} 
where $k$ denotes the suppression factor from the entropy back-reaction, and is expected to be a natural number. The mass loss of PBH is the same with Eq. \ref{eq:massloss_sc}. The slow transition between these two phases, rather than an instantaneous entropy jump, would be affected by a factor $\delta \tau_{\rm SC}/(2t)$, where $\delta$ roughly characterizes the width of the transition region. For the same process, one can integrate Eq. \ref{eq:massloss_mb} to get the mass loss rate, and then the lifetime of a memory-burdened PBH is 
\begin{equation} 
	\tau_{\rm MB}=t_{\rm ev}^k -t_{\rm in}= \tau'_{\rm SC} + \frac{1}{\Gamma_{\rm BH}^k} \simeq (1-q^3)\frac{M_{\rm in}^3}{3\epsilon M_P^4}+\frac{1}{2^k(3+2k)\epsilon M_P}\left(\frac{qM_{\rm in}}{M_P}\right)^{3+2k}. \label{eq:tau_mb} 
\end{equation} 
In this case, a PBH with $M_{\rm in}\gtrsim 2\times 10^{17}{~\rm g}$ can survive until today due to the MB effect for $k=1$ and $q=0.5$. 

\subsection{Neutrino signals} 
Based on the previous section, based on Eq. \ref{eq:emission_rate} and Eq. \ref{eq:massloss_mb}, the neutrino emission rate from a memory-burdened PBH is given by 
\begin{equation} 
	\frac{{\rm d}^2 N_\nu^{\rm MB}}{{\rm d}E_\nu{\rm d}t}=S(M_{\rm PBH})^{-k}\frac{g_\nu}{2 \pi}\frac{\Gamma_{s_\nu}(x)}{e^x+1}. \label{eq:dndedt_mb} 
\end{equation} 
The primary and secondary emission spectra of PBHs and the fragmentation function can be obtained using the \texttt{HDMSpectra} code, which is already integrated into \texttt{BlackHawk 2.30} \cite{Arbey:2019mbc, Arbey:2021mbl}. The observed spectrum is the sum of neutrinos and anti-neutrinos of all flavors, obtained by accounting for decoherent flavor oscillations en route from the PBH to the Earth: 
\begin{equation} 
	\left.\frac{{\rm d}^2 N_{\nu_\alpha}^{\rm MB}}{{\rm d}E_\nu{\rm d}t}\right|_\oplus = \left.\sum_{\beta=e}^\tau P_{\alpha, \beta} \frac{{\rm d}^2 N_{\nu_\beta}^{\rm MB}}{{\rm d}E_\nu{\rm d}t} \right|_{\rm PBH}, 
\end{equation} 
where $P_{\alpha \beta}=\sum_{i=1}^3|U_{\alpha i}|^2|U_{\beta i}|^2$ are the flavour-transition probabilities averaged over cosmological distances to the Earth. The neutrino flavor mixing is governed by the Pontecorvo-Maki-Nakagawa-Sakata (PMNS) mixing matrix $U$ \cite{Pontecorvo:1967fh}. In this work, we adopt the standard three-neutrino framework and use the best-fit mixing angles from global fits: $\theta_{12}=33^\circ$, $\theta_{13}=8.5^\circ$, and $\theta_{23}=45^\circ$ by assuming normal mass ordering \cite{JUNO:2025gmd}. 

A critical aspect for our PBH analysis is that the initial neutrino flux from Hawking radiation is flavor-unbiased, i.e., the production ratio at the source is $1:1:1$. Consequently, the total flux remains equally partitioned among the three flavors after propagation over astronomical distances. Therefore, while the PMNS matrix is essential for predicting flavor-specific signals (e.g., $\nu_\mu$ events at IceCube), the total neutrino flux is independent of the exact values of the mixing angles and CP-violating phases. This implies that the specific choice of $\theta_{23}$ (maximal or non-maximal mixing) actually does not affect the total event rate calculation in this work. Thus, the total dark matter neutrino flux of flavor $\alpha$ detected at the Earth is 
\begin{equation} 
	\frac{{\rm d}\phi_{\nu_\alpha}^{\rm dSph}}{{\rm d}E_\nu {\rm d}\Omega}=\frac{f_{\rm PBH}}{4\pi M_{\rm PBH}} \frac{{\rm d}^2 N_{\nu_\alpha}^{\rm MB}}{{\rm d}E_\nu{\rm d}t} \mathcal{D}(\Omega). 
	\label{eq:flux}
\end{equation} 

\begin{figure}[h] 
	\centering 
	\includegraphics[width=0.49\textwidth]{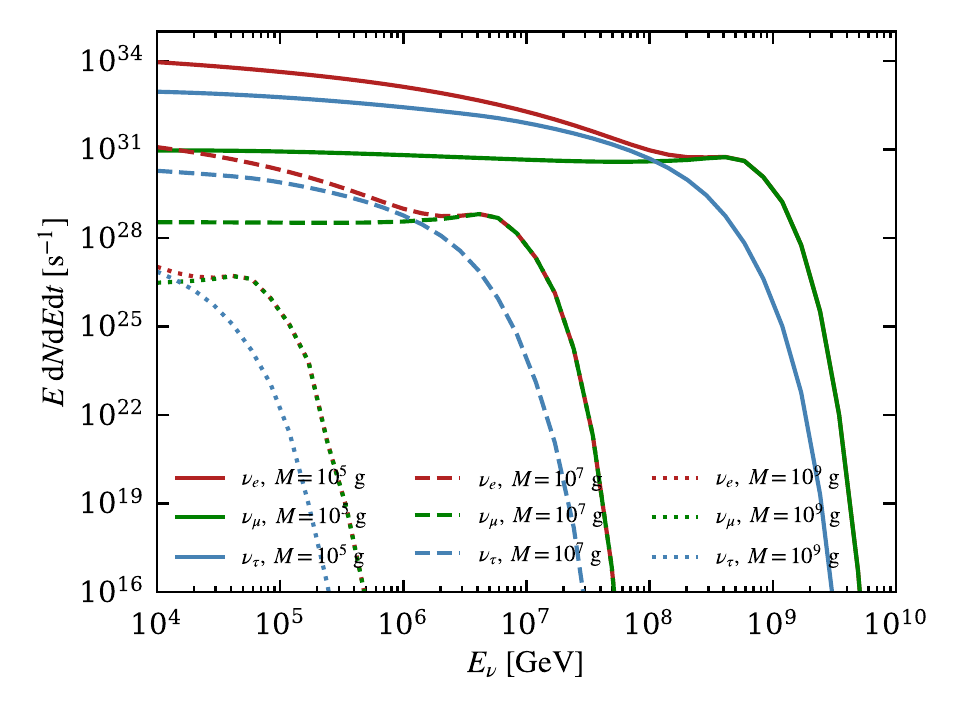} 
	\includegraphics[width=0.49\textwidth]{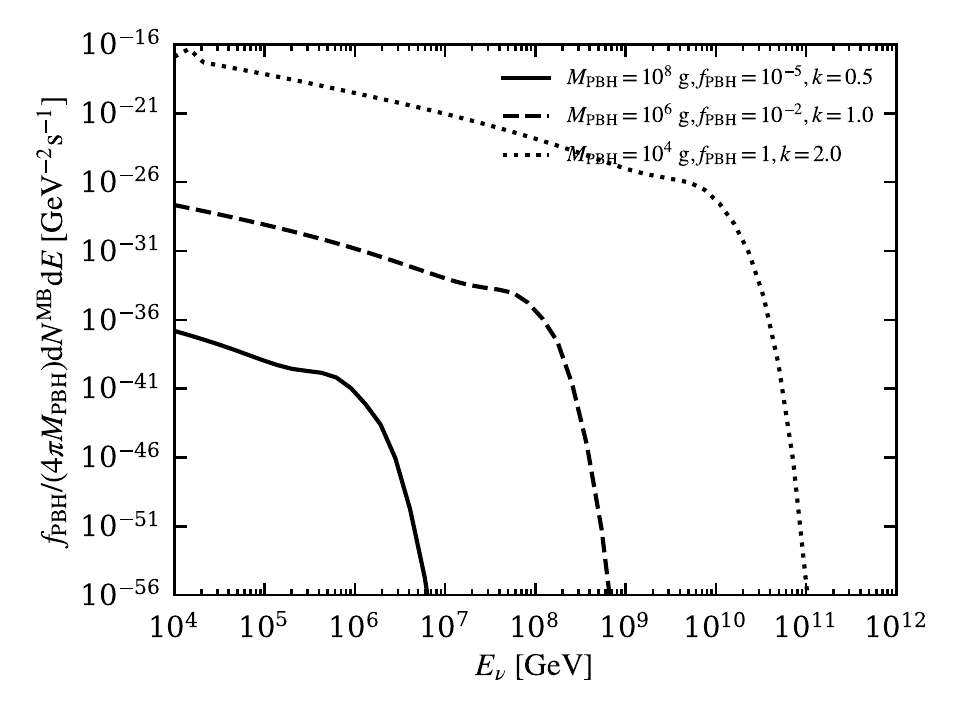} 
	\caption{\textit{Left Panel}: The neutrino number density from memory-burdened PBHs with $k=1$ for different masses $M_{\rm PBH} = 10^{5}$ g, $10^{7}$ g, and $10^{9}$ g. \textit{Right Panel}: The effective number spectra from memory-burdened PBHs without the $\mathcal{D}$-factor in different parameters in MB modes.} \label{fig:dndE_pbhs} 
\end{figure} 

The differential neutrino flux from PBHs in a dSph depends on its DM distribution. The relevant astrophysical quantity is encapsulated by the $\mathcal{D}(\Omega, \theta)$ factor, integrated over the region of interest (ROI) along the line of sight (l.o.s.), which is given by 
\begin{equation} 
	\mathcal{D}(\Omega, \theta)=\frac{1}{\Delta \Omega}\int_{\rm ROI}{\rm d}\Omega \int_{\rm l.o.s} {\rm d}\ell \rho \left( r(\ell,\theta) \right), 
\end{equation} 
where $\Omega$ is the solid angle of the ROI, and $\rho$ represents the dark matter halo profile with radius $r(\ell,\theta)=\sqrt{d^2-2\ell d\cos{\theta}+\ell^2}$ with $d$ the distance of the dSph. 

In this study, we derive constraints on the PBH parameters $f_{\rm PBH}, k, M_{\rm PBH}$, through a combined analysis of multiple dSph observations. A substantial body of literature has been devoted to calculating the astrophysical $\mathcal{D}$- or $\mathcal{J}$-factors that characterize the DM content in these systems. The standard methodology employs the spherical Jeans equation coupled with Bayesian likelihood analysis to constrain the parameters \cite{Evans:2003sc, Geringer-Sameth:2014yza, Bonnivard:2015xpq, Bonnivard:2014kza, Hutten:2016jko, Strigari:2006rd, Strigari_2008, Martinez_2009, Charbonnier:2011ft}. Alternative approaches include establishing scaling relations between the $\mathcal{D}$-factor and observable physical properties of the dSphs \cite{Pace:2018tin}; as well as implementing simple analytic relations that can significantly reduce computational time when transforming $\mathcal{J}$-factor to $\mathcal{D}$-factor values \cite{Evans:2016xwx}. More sophisticated techniques, such as Bayesian hierarchical modeling of the astrophysical properties across the global population of dSphs, have been shown to provide constraints \cite{Martinez:2013els} by leveraging statistical correlations among multiple systems. 

\begin{figure}[h] 
	\centering 
	\includegraphics[width=0.49\textwidth]{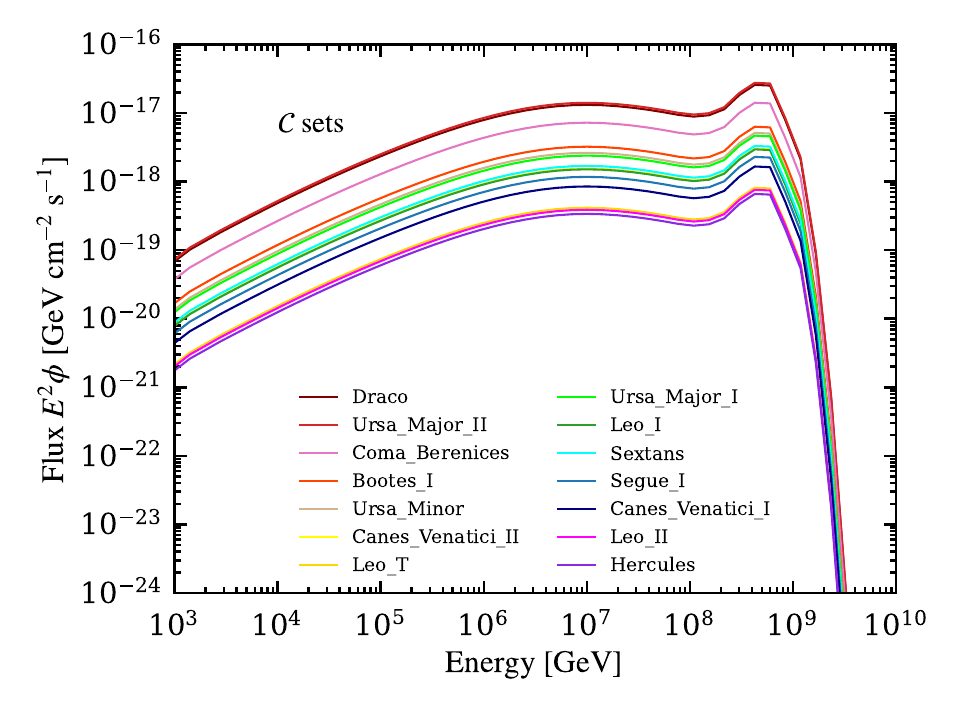} 
	\includegraphics[width=0.49\textwidth]{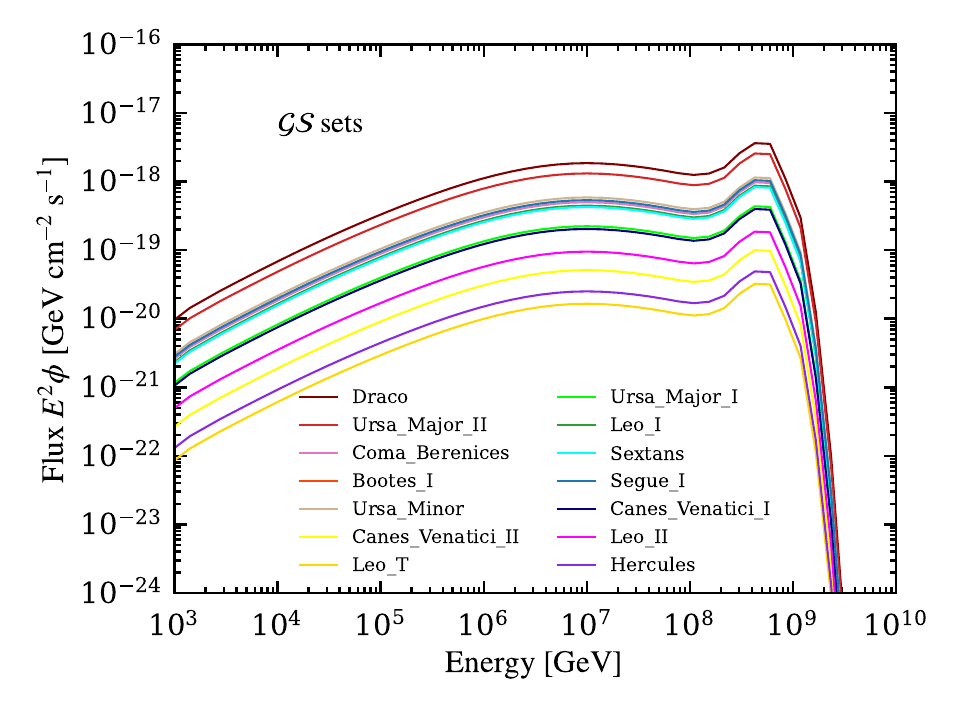} 
	\includegraphics[width=0.49\textwidth]{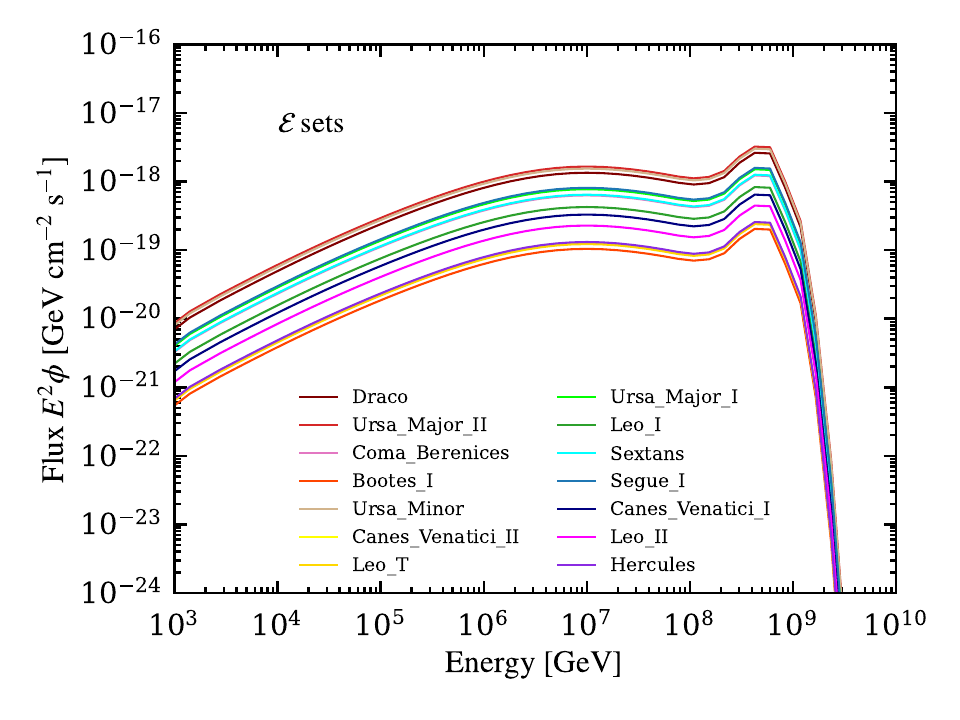} 
	\includegraphics[width=0.49\textwidth]{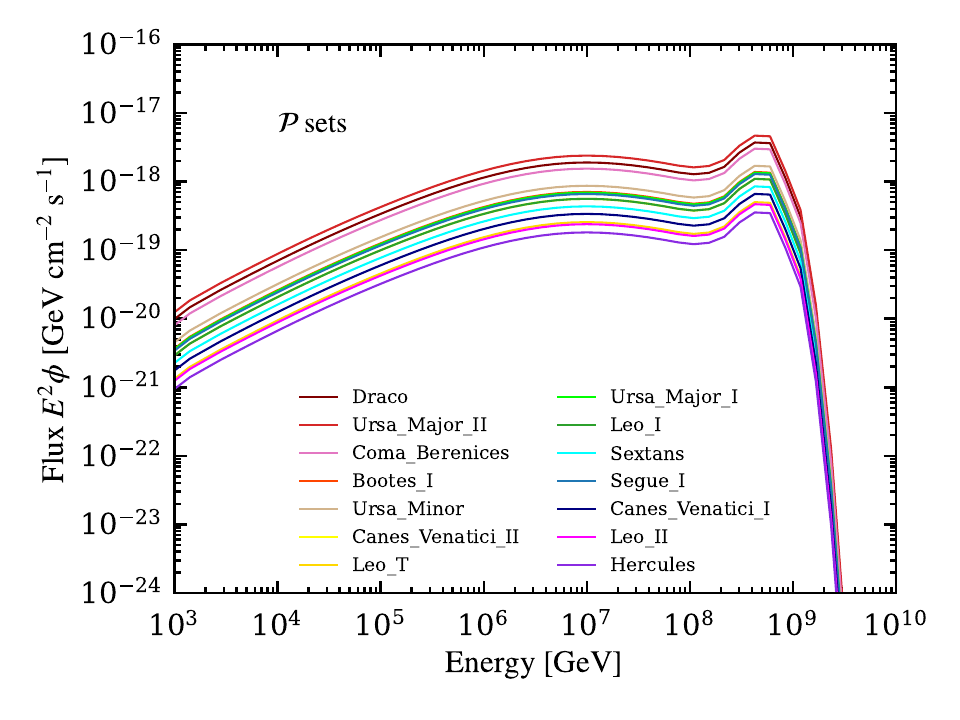} 
	\caption{Comparison of neutrino fluxes from memory-burdened PBHs for different dSphs and their source datasets. The parameters are set to $M_{\rm PBH} = 10^{5}$ g, $k=2$, and $f_{\rm PBH}=10^{-5}$. The different datasets are labeled in each figure.} \label{fig:flux_pbhs} 
\end{figure} 

The continuous discovery of new dSphs by current observational facilities has significantly enhanced their importance for dark matter searches. As the catalog of known dSphs continues to expand, obtaining reliable $\mathcal{D}$-factor estimates becomes increasingly crucial for robust indirect detection analyses. Consequently, in this work, we have assembled a comprehensive sample of classical dSphs from various methodological approaches, like in ref. \cite{Fermi-LAT:2025gei}. Below, we outline the rationale for our selection. 

For the standard computational methodology involving detailed Jeans models analyzed through Markov Chain Monte Carlo techniques, we adopt the results from Geringer-Sameth et al. \cite{Geringer-Sameth_2015} (hereafter referred to as the $\mathcal{GS}$ set). This study employed a uniform analysis framework applied to the stellar-kinematic data available for 20 Milky Way dSphs at the time of their analysis. Additionally, we incorporate the simple analytic formulae developed by Evans et al. \cite{Evans:2016xwx} (or $\mathcal{E}$ set), which offers significant computational efficiency advantages over traditional Jeans solution methods while maintaining reasonable accuracy. This approach enables valuable comparisons between different computational strategies. Furthermore, we include the results from Pace \& Stigari \cite{Pace:2018tin} (or $\mathcal{P}$ set), which derives $\mathcal{D}$-factor estimates through empirical scaling relations connecting the dark matter content to observable physical properties of the dSphs, including velocity dispersion, distance, and stellar half-light radius. This method bypasses the need for full dynamical modeling while still providing physically motivated estimates. The $\mathcal{C}$ set incorporates recent observations from the Large High Altitude Air Shower Observatory (LHAASO) \cite{LHAASO:2024upb}, where the collaboration has updated measurements for 16 dSphs within their field-of-view (FoV). These observations provide improved constraints, particularly for heavier dark matter candidates. 

The detailed information for all dSphs used in our PBH analysis is presented in Tab. \ref{tab:DSphs_info}. It is important to note that the different datasets contain varying numbers of dSphs. To ensure consistency across methodologies, we selected only those dSphs that appear in all four datasets, resulting in a final sample of 14 dSphs. Additionally, we observe that the $\mathcal{C}$ set adopts the angle between the center of the dwarf and an estimate of the distance to the outermost member star, whereas the other sets used in this analysis consistently employ $0.5^\circ$ as their angular cut. This methodological difference likely explains why the $\mathcal{D}$-factor values from the $\mathcal{C}$ set are generally higher than those from other datasets. Under these adoptions of $\mathcal{D}$-factors, Fig. \ref{fig:flux_pbhs} displays the neutrino flux from memory-burdened PBHs for distinct dSphs. 

\begin{table*} 
	\centering 
	\setlength{\tabcolsep}{5pt} 
	\caption{Comprehensive information on the dSphs with their $\mathcal{D}$-factors considered in the present work. Column 1 lists the names of the dSphs, while their coordinates and heliocentric distances are provided in columns 2 and 3, respectively. The remaining columns present the $\mathcal{D}$-factors collected from the independent $\mathcal{C}, \mathcal{GS}, \mathcal{E}$ and $\mathcal{P}$ sets, along with their corresponding $\pm1 \sigma$ uncertainties. To enable a consistent combination of all dSph data, we selected only those sources that appear in all four datasets. The top section contains the classical dwarf galaxies, whereas the bottom section displays the ultrafaint dwarf galaxies.} 
	\begin{tabular}{ccccccccc} 
		\hline \hline 
		DSphs & RA & DEC & $d$ & $r_{1/2}$& $\log_{10}\mathcal{D}$ & & & \\ 
		(name)& (deg) & (deg) & (kpc) & (pc) & ($\rm GeV~cm^{-2}$)&&& \\ 
		& & & & &($\mathcal{C}$ set) & ($\mathcal{GS}$ set) & ($\mathcal{E}$ set) & ($\mathcal{P}$ set) \\ 
		\hline 
		Draco & 260.05 & 57.92 & 82 & 182 & $19.38^{+0.24}_{-0.32}$ & $18.53^{+0.10}_{-0.12}$ & $18.39^{+0.25}_{-0.25}$ & $18.54^{+0.11}_{-0.14}$ \\ 
		Leo I & 152.12 & 12.30 & 250 & 292 & $18.44^{+0.33}_{-0.42}$ & $17.91^{+0.15}_{-0.20}$ & $17.89^{+0.28}_{-0.28}$ & $18.01^{+0.20}_{-0.28}$ \\ 
		Leo II & 168.37 & 22.15 & 205 & 158 & $17.85^{+0.62}_{-0.40}$ & $17.24^{+0.35}_{-0.48}$ & $17.62^{+0.25}_{-0.25}$ & $17.64^{+0.50}_{-0.33}$ \\ 
		Sextans & 153.26 & -1.61 & 86 & 523 & $18.49^{+0.28}_{-0.21}$ & $17.89^{+0.13}_{-0.23}$ & $18.07^{+0.29}_{-0.29}$ & $17.90^{+0.11}_{-0.09}$ \\ 
		Ursa Minor & 227.28 & 67.23 & 66 & 269 & $18.68^{+0.33}_{-0.15}$ & $18.03^{+0.16}_{-0.13}$ & $18.45^{+0.24}_{-0.24}$ & $18.20^{+0.14}_{-0.08}$ \\ 
		\hline 
		Bo\"otes I & 210.02 & 14.50 & 66 & 187 & $18.77^{+0.40}_{-0.54}$ & $17.90^{+0.23}_{-0.26}$ & $17.28^{+0.64}_{-0.38}$ & $18.11^{+0.25}_{-0.30}$ \\ 
		Canes Venatici I & 202.02 & 33.56 & 218 & 423& $18.19^{+0.40}_{-0.39}$ & $17.57^{+0.36}_{-0.72}$ & $17.78^{+0.11}_{-0.11}$ & $17.79^{+0.26}_{-0.27}$ \\ 
		Canes Venatici II & 194.29 & 34.32 & 160 & 68 & $18.45^{+0.50}_{-0.74}$ & $16.97^{+0.24}_{-0.23}$ & $17.37^{+0.40}_{-0.40}$ & $18.01^{+0.36}_{-0.51}$ \\ 
		Coma Berenices & 186.74 & 23.90 & 44 & 57 & $19.12^{+0.46}_{-0.73}$ & $17.96^{+0.20}_{-0.25}$ & $18.06^{+0.32}_{-0.32}$ & $18.45^{+0.29}_{-0.44}$ \\ 
		Hercules & 247.76 & 12.79 & 132 & 98 & $17.79^{+0.62}_{-0.61}$ & $16.66^{+0.42}_{-0.40}$ & $17.38^{+0.45}_{-0.45}$ & $17.52^{+0.50}_{-0.51}$ \\ 
		Leo T & 143.72 & 17.05 & 407 & 142 & $17.88^{+0.65}_{-0.69}$ & $16.48^{+0.22}_{-0.25}$ & $17.35^{+0.37}_{-0.37}$ & $17.67^{+0.53}_{-0.60}$ \\ 
		Segue I & 151.77 & 16.08 & 23 & 21 & $18.33^{+0.69}_{-0.63}$ & $17.99^{+0.20}_{-0.31}$ & $18.17^{+0.39}_{-0.39}$ & $18.08^{+0.53}_{-0.49}$ \\ 
		Ursa Major I & 158.71 & 51.92 & 97 & 199 & $18.64^{+0.50}_{-0.48}$ & $17.61^{+0.20}_{-0.38}$ & $18.15^{+0.25}_{-0.25}$ & $18.10^{+0.28}_{-0.29}$ \\ 
		Ursa Major II & 132.87 & 63.13 & 30 & 99 & $19.41^{+0.43}_{-0.57}$ & $18.38^{+0.25}_{-0.27}$ & $18.48^{+0.39}_{-0.39}$ & $18.64^{+0.28}_{-0.31}$ \\ 
		\hline \hline 
	\end{tabular} 
	\label{tab:DSphs_info} 
\end{table*} 

\section{Data Analysis}\label{sec:ana} 
In this section, then we analyze the observational data from IceCube and construct the likelihood function to determine the possibility that any high-energy neutrino signals are caused by memory-burdened PBHs in those 14 dSphs described above. 

\subsection{IceCube Data} 
The IceCube Neutrino Observatory represents a monumental achievement in high-energy astroparticle physics, transforming a cubic kilometer of pristine Antarctic ice into the world’s largest neutrino detector. Located at the geographic South Pole, this sophisticated instrument consists of 5,160 digital optical modules arranged along 86 vertical strings, strategically deployed at depths ranging from 1,450 to 2,450 meters beneath the ice surface \cite{Aartsen_2017}. While originally designed to detect astrophysical neutrinos across the 100 TeV to several EeV energy range, IceCube’s unique configuration and unprecedented sensitivity make it exceptionally well-suited for probing exotic phenomena such as neutrino emission from PBHs. 

For our investigation, we utilize the complete decade-long dataset of muon-track events publicly released by the IceCube Collaboration, comprising 1,134,450 carefully selected events. This comprehensive dataset, organized into ten distinct operational configurations (designated IC40, IC59, IC79, and IC86-I through IC86-VII, where the numerical identifier reflects the number of active detector strings), provides the temporal coverage and statistical power essential for our search for weak signals from distant dSphs. The publicly available data package \cite{IceCube:2021xar} encompasses all necessary components for our analysis: the experimental event records, detector operational statistics, energy distributions, instrumental response matrices, and effective areas across different detector configurations. 

\subsection{Likelihood function}
We try to quantify the preference of the IceCube data for PBH origin, here we adopt a likelihood-ratio test where the the test statistic (TS) defined as 
\begin{equation} 
	{\rm TS}=-2 \ln{\left(\frac{\mathcal{L}(\hat{\mu}, \hat{\boldsymbol{\xi}})}{\mathcal{L}(\mu_0, \hat{\boldsymbol{\xi}})} \right)}, 
	\label{eq:TS}
\end{equation} 
where $\hat\mu$ are the parameters of the PBH model, which contains $f_{\rm PBH}$ and $k$, and $\hat{\boldsymbol{\xi}}$ is the set of nuisance parameters that includes both the neutrino background, IceCube event characteristics, and the dSph $\mathcal{D}$-factor from Tab. \ref{tab:DSphs_info}. We define the \textit{total} joint likelihood function, which describes all observations from the detector's 3D unbinned likelihood and incorporates dSphs, by factoring it into \textit{partial} joint likelihood functions corresponding to each different $\mathcal{D}$-factor: 
\begin{equation} 
	\begin{aligned} 
		\mathcal{L}(\hat{\mu}, \hat{\boldsymbol{\xi}}) 
		&= \max_{\mathcal D}\prod_{l=1}^{N_l} \prod_{j}\prod_{i=1}^{N_j} \left[ \mathcal{L}_i^j(\hat{\mu}^j, \hat{\boldsymbol{\xi}}_i^j|\mathcal{D}_l) \times \mathcal{L}_{\mathcal{D}}(\mathcal{D}_l|\mathcal{D}_{0,l},\sigma_{\mathcal{D}_l}) \right] \\ 
		&=\max_{\mathcal D}\prod_{l=1}^{N_l} \prod_{j}\prod_{i=1}^{N_j} \left[ \left( \frac{\hat{\mu}^j \mathcal{D}_l}{N_j} S_i^j + \left( 1-\frac{\hat{\mu}^j \mathcal{D}_l}{N_j} \right) B_i^j \right)\times \mathcal{L}(\mathcal{D}_l|\mathcal{D}_{0,l},\sigma_{\mathcal{D}_l})\right],\\ 
	\end{aligned} 
\end{equation} 
where $\mathcal{L}(\mathcal{D}_l|\mathcal{D}_{0,l},\sigma_{\log \mathcal{D}_l})$ is the likelihood term for the $\mathcal{D}$-factor, and $l$ runs over $N_l=14$ denoting the total number of dSphs. For the $\mathcal{D}$-factor, we adopt an asymmetric Gaussian prior to account for the asymmetric uncertainties reported in the literature. It is defined as 
\begin{equation} 
	\mathcal{L}_{\mathcal{D}}(\mathcal{D}_l|\mathcal{D}_{0,l},\sigma_{\mathcal{D}_l}) = \frac{1}{\ln(10) \mathcal{D}_{0,l}\sqrt{2\pi}\sigma_{ \mathcal{D}_l}} \exp{\left[-\frac{(\mathcal{D}_l-\mathcal{D}_{0,l})^2}{2 \sigma^2_{\mathcal{D}_l}} \right]}. \label{eq:likelihood_dfactor} 
\end{equation} 
where $\mathcal{D}_{0,l}$ refers to the measured value for null hypothesis. We preserve the original asymmetry of the measurement uncertainties in error $\sigma_{\mathcal{D}_l}$. The width of the Gaussian is chosen conditionally based on the sign of the deviation: $\sigma_{\rm low}$ is used when $\mathcal{D}<\mathcal{D}_0$, and $\sigma_{\rm high}$ otherwise. In the profile likelihood maximization over the nuisance parameter $\mathcal{D}$, we employ an asymmetric scanning range of $[D_0-5\sigma_{\rm low}, \mathcal{D}_0+5\sigma_{\rm high}]$ to fully cover the posterior region.. Both sets in Tab. \ref{tab:DSphs_info} will be used independently in our analysis. The right hand of Eq. \ref{eq:likelihood_dfactor} can be interpreted both as the likelihood function and as the probability density function (PDF) for the associated random variable $\mathcal{D}_{0,l}$ when it is normalized. 

To determine whether IceCube has detected neutrinos emitted from PBHs, we define a likelihood $\mathcal{L}$ for observing $n_s$ signal events from a source, given by the product of probability density functions (PDFs) for each $i$-th muon-track event in the $j$-th data sample. We then analyze the neutrino emission from dSphs, as given by the first term in parentheses in the second line of the Eq. \ref{eq:likelihood_dfactor}. Here $N_j$ is the total number of events within the region of interest (ROI), defined as a $5^\circ$ circular region around the target source, and $\hat{\mu}^j \mathcal{D}_l$ represents the number of signal events from the sources, so 
\begin{equation} 
	\hat{\mu}^j = \kappa 4 \pi T^j_{\rm obs} \int_{E_{\rm min}}^{E_{\rm max}} {\rm d}E_\nu A_{\rm eff}^j(E_\nu, \delta_s) \left.\frac{{\rm d}^2 N_{\nu_\alpha}^{\rm MB}}{{\rm d}E_\nu{\rm d}t}\right|_\oplus, 
\end{equation} 
with effective area $A_{\rm eff}^j$, and $\kappa=f_{\rm PBH}/(4\pi M_{\rm PBH})$. This depends on the neutrino energy $E_\nu$ and the declination angle $\delta_s$. $T^j_{\rm obs}$ is the total observation time of sample $j$, recording the time period during which the detector was operational. The $S_i^j$ and $B_i^j$ are the PDFs for signal and background events for the $j$-th data sample, given by 
\begin{equation} 
	S_i^j = S_{\rm spat}^j (\vec{x}_i| \sigma_i, \vec{x}_s, \sigma_s) S^j_{\rm ener}(E_i|\vec{x}_i, \gamma), 
\end{equation} 
\begin{equation} 
	B_i^j = B_{\rm spat}^j(\delta_i) B_{\rm ener}^j(E_i|\delta_i). 
\end{equation} 
Here, $S_{\rm spat}^j$ is the spatial signal PDF that describes the distribution of the reconstructed direction of signal events $\vec{x}_i$. It follows a 2-dimensional Gaussian distribution with source extension $\sigma_s$ and event angular uncertainty $\sigma_i$, which can be written as 
\begin{equation} 
	S_{\rm spat}^j (\vec{x}_i| \sigma_i, \vec{x}_s, \sigma_s)=\frac{1}{2\pi (\sigma_s^2 + \sigma_i^2)} \exp{ \left[ -\frac{D(\vec{x}_i, \vec{x}_s)^2}{2(\sigma_s^2 + \sigma_i^2)} \right] \frac{D(\vec{x}_i, \vec{x}_s)}{\sin{D(\vec{x}_i, \vec{x}_s)}}}, 
\end{equation} 
where $D(\vec{x}_i, \vec{x}_s)$ represents the angular distance between the true and reconstructed directions, the spatial signal PDF is normalized as $\int S_{\rm spat}^j (\vec{x}_i| \sigma_i, \vec{x}_s, \sigma_s) {\rm d}\Omega = 1$. Typically, $\sigma_i \approx 0^\circ.64$ for track-like events \cite{Halzen:2016seh}. Then, the energy signal PDF $S_{\rm ener}^j$ describes the distribution of the reconstructed energy of signal events, given by 
\begin{equation} 
	S^j_{\rm ener}(E_i|\vec{x}_i, \gamma) = \frac{ \int \Phi_\nu(E_\nu)A^j_{\rm eff}(E_\nu, \delta_s)\mathcal{M}_j(E_i|E_\nu, \delta_s){\rm d}E_\nu}{\int \Phi_\nu(E_\nu)A_{\rm eff}^j(E_\nu, \delta_s){\rm d}E_\nu}, 
\end{equation} 
where $\mathcal{M}_j$ is the smearing function. This gives the fractional count of simulated events in the reconstructed energy bin relative to all events in the $(E_\nu, \delta_s)$ bin, representing the probability of obtaining the reconstructed energy $E_{\rm rec}$ when a neutrino with energy $E_\nu$ enters the detector at declination $\delta_s$. Finally, the background spatial PDF and the energy term of the background PDF are 
\begin{equation} 
	B_{\rm spat}^j(\delta_i)=\frac{N_{\delta_i \pm 3}^j}{N_j \times \Delta \Omega}, 
\end{equation} 
and 
\begin{equation} 
	B_{\rm ener}^j(E_{\rm rec}|\delta_i) = \frac{N_{i'j}^j}{N_{\delta_i \pm 3}^j \Delta E_{i'}}, 
\end{equation} 
where $N_{\delta_i \pm 3}^j$ is the number of events within a ring region of $\delta\pm 3^\circ$ and $\Delta \Omega = 2\pi \Delta \sin{\delta}$ is its solid angle. For data sample $j$, $N_{i'j}^j$ is the number of events with the declination $\sin{\delta} \in [\sin{\delta_i}, \sin{\delta_i}+0.02]$ and the reconstructed energy range of $\log_{10}{E_{\rm rec}} \in [\log_{10}{E_{i'}}, \log_{10}{E_{i'}}+0.1]$. 

We implement the likelihood and found no excess signal over the background-only hypothesis, so we derived the 95\% confidence level (C.L.) upper bounds for PBH parameters by setting Eq. \ref{eq:TS} to a value of 2.71.

\begin{figure}[h] 
	\centering 
	\includegraphics[width=0.49\textwidth]{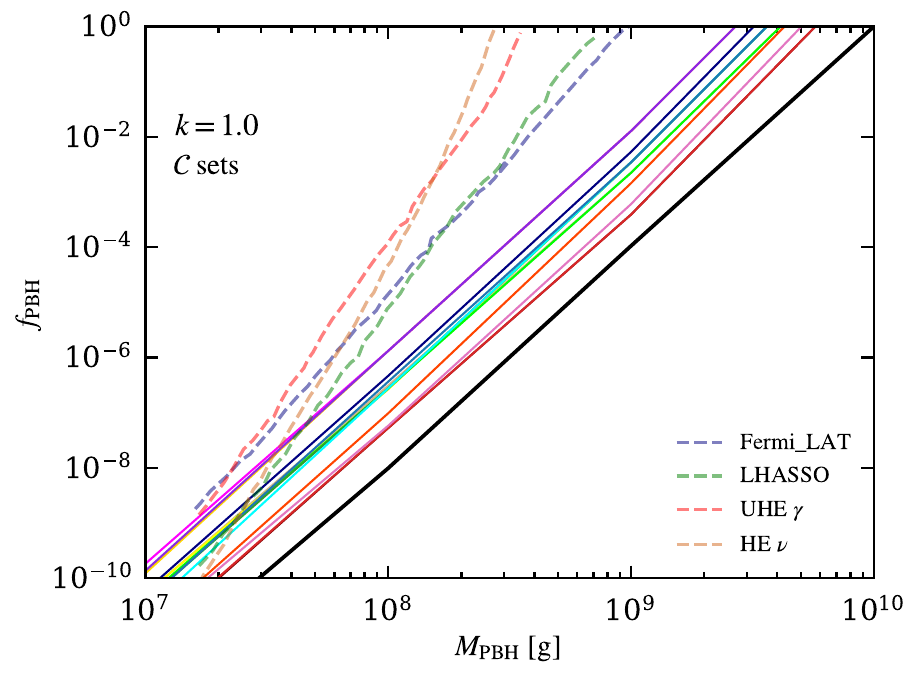} 
	\includegraphics[width=0.49\textwidth]{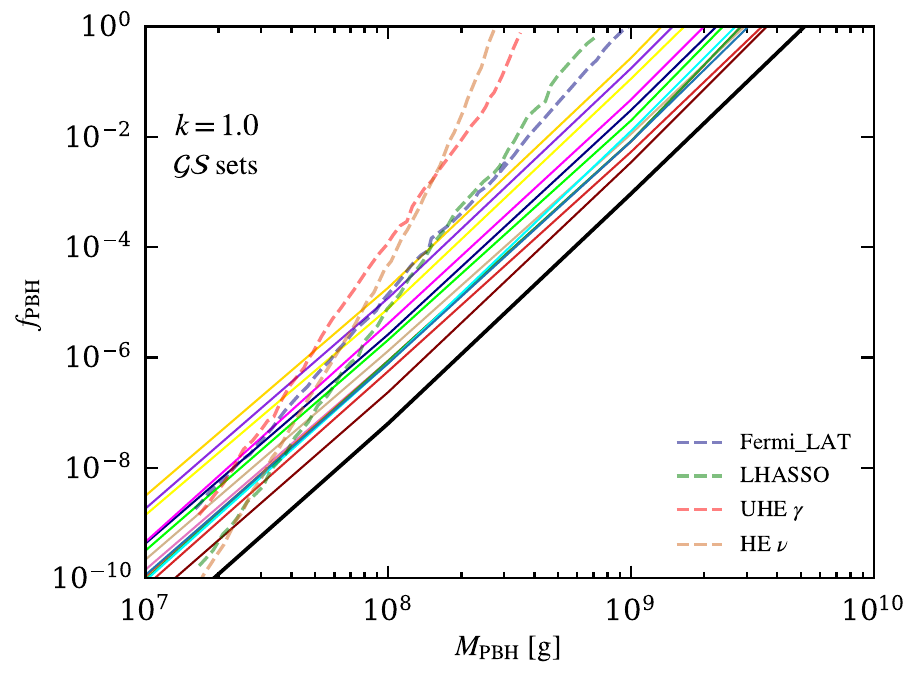} 
	\includegraphics[width=0.49\textwidth]{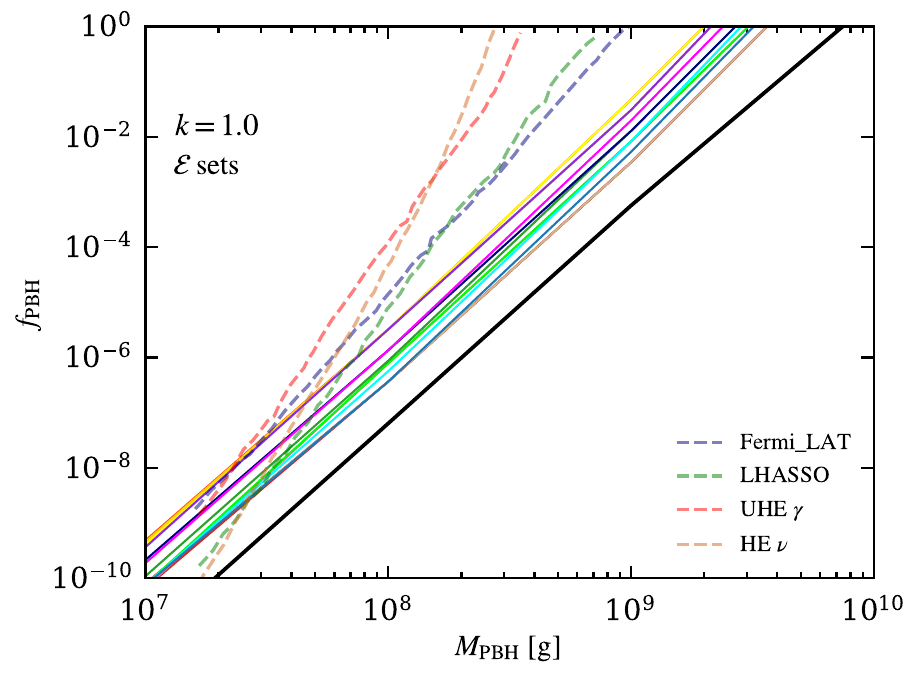} 
	\includegraphics[width=0.49\textwidth]{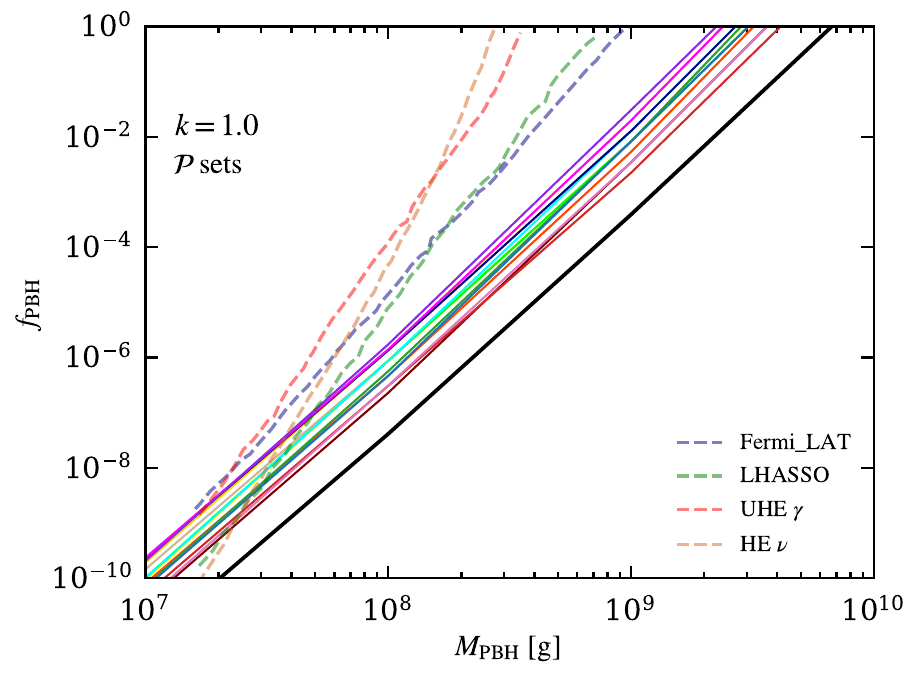} 
	\caption{The 95\% C.L. upper limits on the $f_{\rm PBH}-M_{\rm PBH}$ plane derived from four dSph datasets for $k=1$. Previous results are shown as dashed lines for comparison. The black solid line represents the combined constraint from all dSphs, which is significantly tighter than the limits from individual dSphs. 
		The color legend for the dSphs datasets is identical to that in the Fig. \ref{fig:flux_pbhs}, see text for more details.}
	\label{fig:flimits_k1} 
\end{figure} 

\begin{figure}[h] 
	\centering 
	\includegraphics[width=0.49\textwidth]{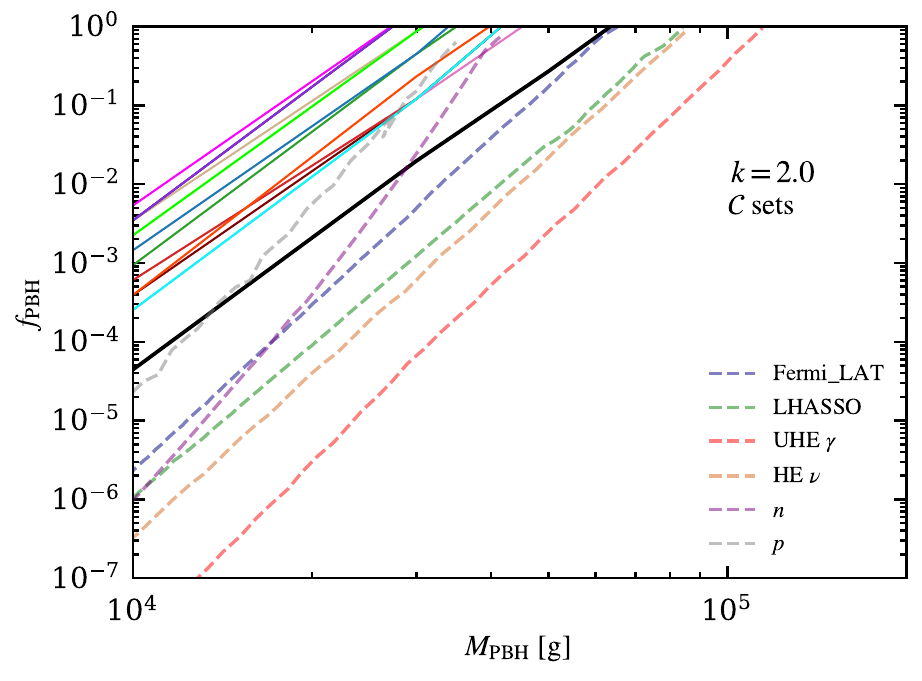} 
	\includegraphics[width=0.49\textwidth]{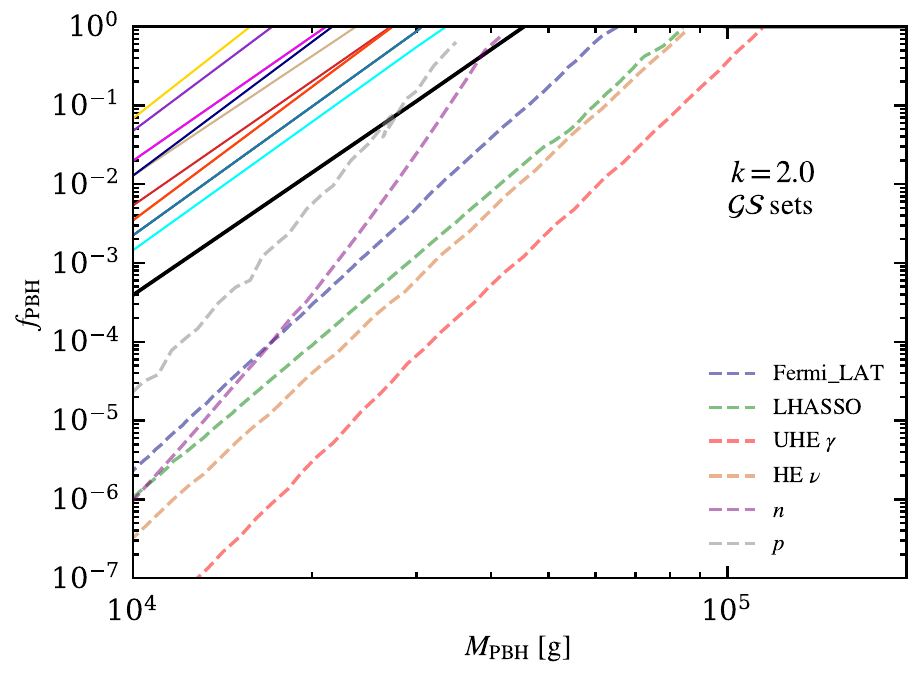} 
	\includegraphics[width=0.49\textwidth]{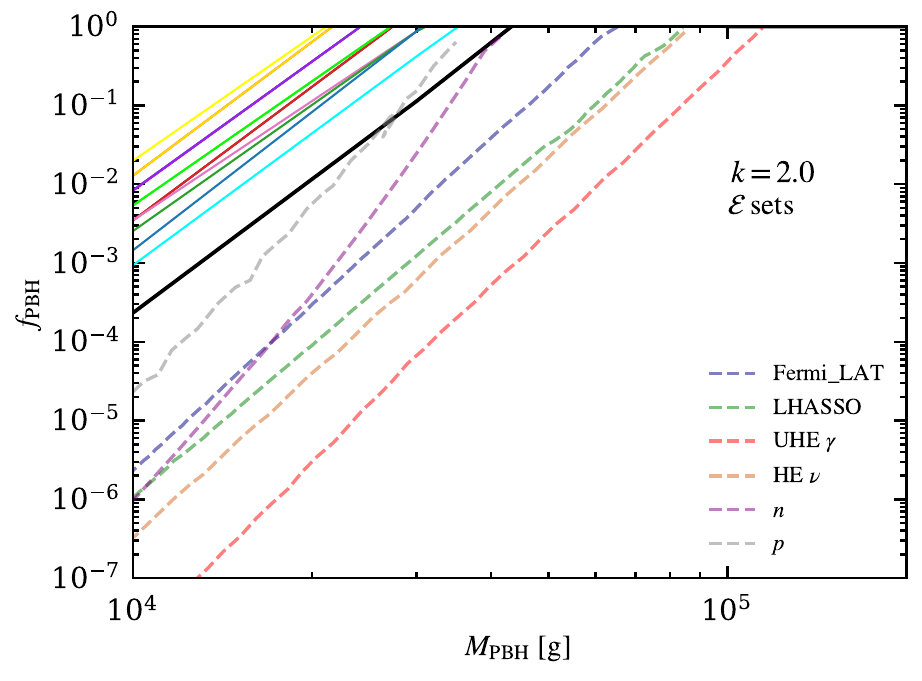} 
	\includegraphics[width=0.49\textwidth]{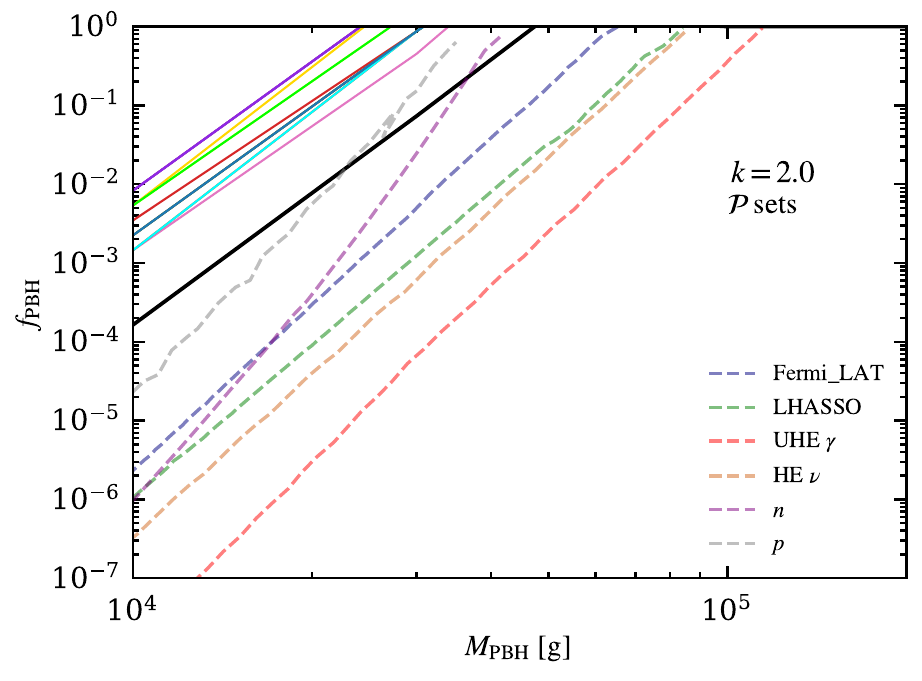} 
	\caption{The 95\% C.L. upper limits on the $f_{\rm PBH}-M_{\rm PBH}$ plane derived from four dSph datasets for $k=2$. Previous results are shown as dashed lines for comparison. The black solid line represents the combined constraint from all dSphs, which is significantly tighter than the limits from individual dSphs. The color legend for the dSphs datasets is identical to that in the Fig. \ref{fig:flux_pbhs}, see text for more details.}
	\label{fig:flimits_k2} 
\end{figure} 

\section{Results and Discussion}\label{sec:res} 
\begin{figure}[h] 
	\centering 
	\includegraphics[width=0.49\textwidth]{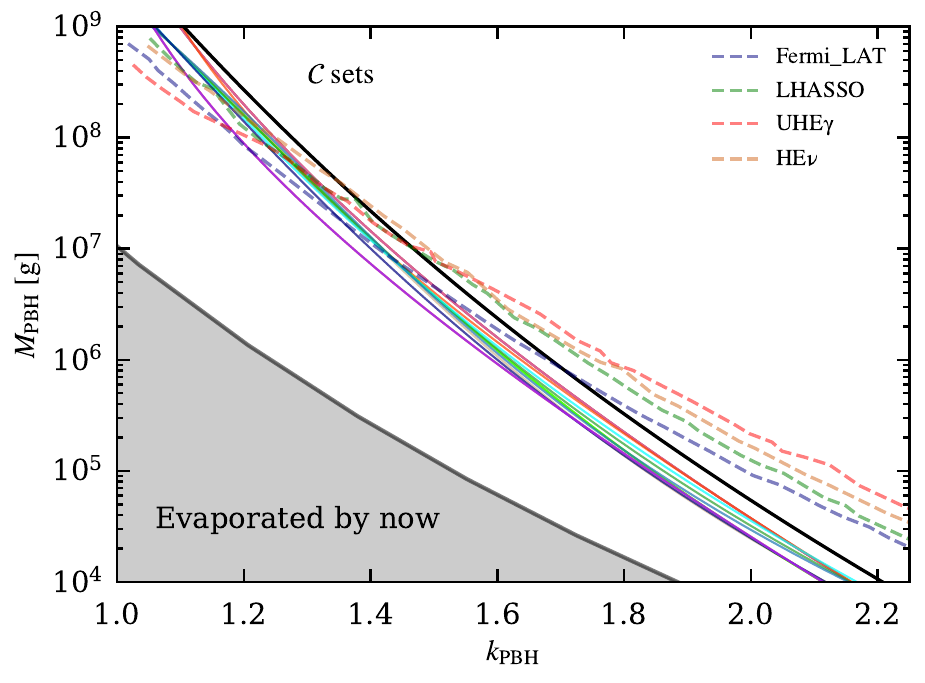} 
	\includegraphics[width=0.49\textwidth]{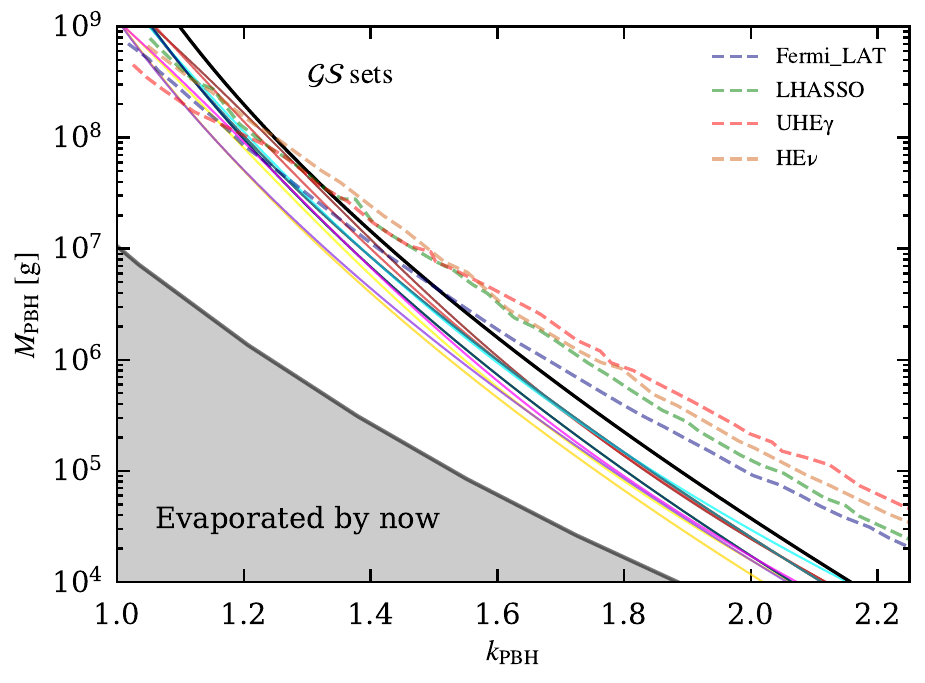} 
	\includegraphics[width=0.49\textwidth]{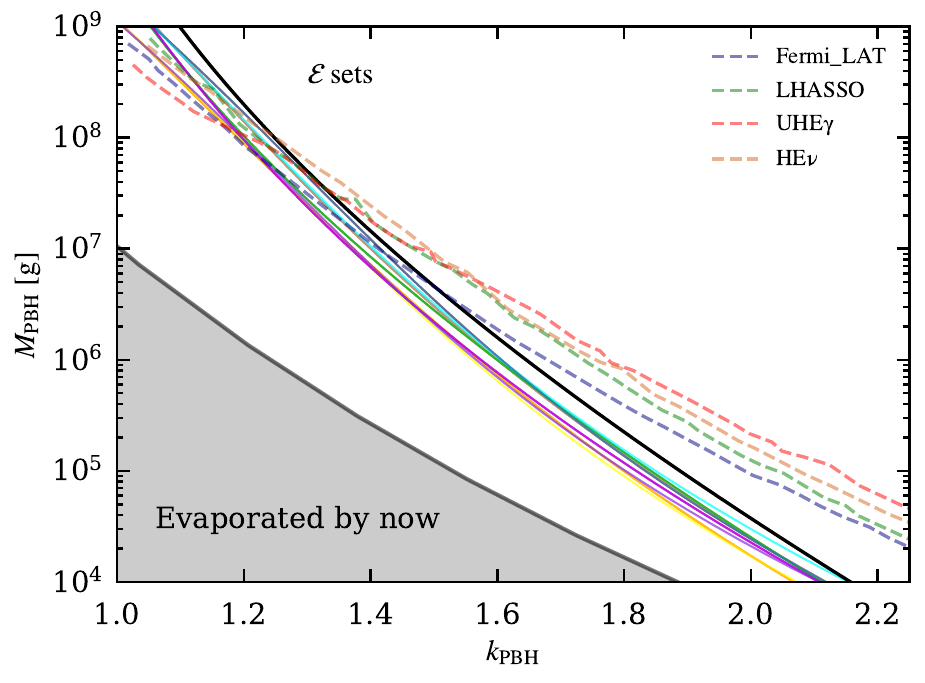} 
	\includegraphics[width=0.49\textwidth]{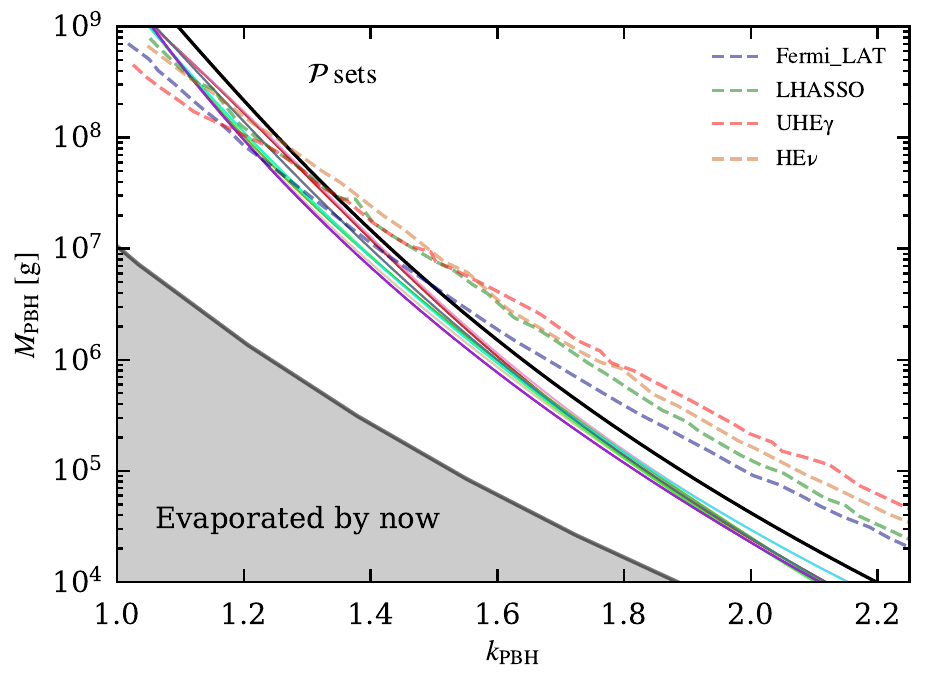} 
 	\caption{The 95\% C.L. upper limits on the $k-M_{\rm PBH}$ plane derived from four dSph datasets for $f_{\rm PBH}=1$. Previous results are shown as dashed lines for comparison. The black solid line represents the combined constraint from all dSphs, which is significantly tighter than the limits from individual dSphs. The color legend for the dSphs datasets is identical to that in the Fig. \ref{fig:flux_pbhs}, see text for more details.}
	\label{fig:klimits} 
\end{figure}

We present and analyze our results in this section. It worth noting that the previous constraints compared in our figures were derived under the assumption of mass tracking. However, according to Ref. \cite{Dondarini:2025ktz} and Eq. \ref{eq:flux}, PBHs are no longer self-similar after the onset of the MB effect. Therefore, we have applied appropriate shifts to the cited results to ensure a valid comparison.
Fig. \ref{fig:flimits_k1} and Fig. \ref{fig:flimits_k2} display the constraints on PBH abundance $f_{\rm PBH}$ for $k=1$ and $k=2$ cases, respectively. The relation between $k$ and $M_{\rm PBH}$ is shown in Fig. \ref{fig:klimits}. The different line colors represent limits obtained from distinct datasets: the dashed navy line, dashed green line and dashed red line are from Fermi-LAT, LHAASO, ultra-high-energy diffuse gamma-ray observations \cite{Chianese:2025wrk}, respectively; the dashed brown line for high-energy neutrinos is taken from Ref. \cite{Chianese:2024rsn}, dashed purple and gray lines are from cosmic-ray (proton and neutron, respectively) \cite{Ambrosone:2026djo}.
Our findings reveal several key insights: 
\begin{enumerate} 
	\item Generally, the $\mathcal{D}$-factor values play a crucial role in determining the strength of the constraints, as larger $\mathcal{D}$-factor values correspond to higher dark matter densities, which naturally produce stronger neutrino fluxes and consequently more detectable events. This highlights the importance of precise $\mathcal{D}$-factor measurements and justifies our use of multiple $\mathcal{D}$-factor sets for a comparative analysis. This explains why the $\mathcal{C}$ set delivers superior constraints in both $k$ value cases. However, the spatial configuration of individual dSphs also significantly impacts the analysis effectiveness. The spatial probability distribution functions between dSph locations and IceCube neutrino events must be carefully considered, as sources positioned far from most detected events inevitably yield weaker constraints. This suggests that future analyses with higher event statistics and improved spatial coverage could yield substantially enhanced constraints. 
	
	\item The $k=1$ scenario is significantly more advantageous than the $k=2$ case. Even under the conservative lower $\mathcal{D}$-factor set like $\mathcal{GS}$ and $\mathcal{E}$ set, still yielding significantly tighter constraints than those reported in previous studies.
	This superiority stems from fundamental differences in the spectral characteristics. As shown in the left panel of Fig.~\ref{fig:dndE_pbhs}, the spectra derived from lighter PBHs dominate higher energy emissions. When considering IceCube's optimal energy range and effective areas, PBH masses produce markedly different responses. The spectral peaks of PBHs with specific $M_{\rm PBH}$ values align particularly well with the detector's sensitivity range. Under the MB effect, the parameter $k$ influences the strength of the entropy suppression, which can substantially reduce the flux and define the viable parameter space for PBHs. For $k=2$, as evidenced in Eq.~\ref{eq:dndedt_mb} and the right panel of Fig.~\ref{fig:dndE_pbhs}, the strong suppression due to entropy effects significantly limits detectable events. In the $k=1$ case, PBHs with masses ranging from $10^7$ to $10^{10}$~g contribute appreciable event numbers, while the $k=2$ case shows severe suppression. Remarkably, for the $k=1$ case, the constraints derived in this work can exclude PBHs as the dominant component of dark matter up to approximately $10^{10}$~g, which significantly improves upon previous results. With higher energy neutrino data coverage, potentially achievable with future instruments like IceCube-Gen2~\cite{IceCube-Gen2:2020qha}, the $k=2$ case could yield competitive constraints.
	
	\item The $k-M_{\rm PBH}$ relation plane shown in Fig. \ref{fig:klimits} provides additional insights into the MB effect. The grey filled region indicates PBH masses that would have completely evaporated by the present epoch under different $k$ values. While our constraints on lighter PBHs are currently much less stringent than existing results, we observe a clear trend with a different slope compared to previous studies, where our analysis becomes increasingly competitive for higher PBH masses. This behavior also reflects the growing sensitivity of the IceCube detector to harder neutrino spectra characteristic of more massive PBHs under the MB framework. 
	
	\item To ensure clarity, the dSph legend conventions are kept consistent across Fig. \ref{fig:flimits_k1}, Fig. \ref{fig:flimits_k2}, and Fig. \ref{fig:klimits}. Among the 14 dSphs, Ursa Major II provides the strongest constraints for $k=1$, owing to its high $\mathcal{D}$-factor value. For $k=2$, the expected neutrino signals from all dSphs are too weak to yield meaningful constraints. Notably, the black curves represent the combined analysis incorporating all 14 dSphs. This joint analysis, which yields the final exclusion limits of this work, clearly produces substantially improved limits over the entire parameter space, highlighting the statistical power of combining multiple targets with independent systematic uncertainties. Our results robustly demonstrate that the neutrino emission from PBHs in dSphs yields stronger constraints than previous studies for $k\lesssim 1.3$, and even for $k\lesssim 1.5$ when all dSphs are combined.

\end{enumerate} 

\section{Conclusion}\label{sec:concl} 
In this work, we investigate possible neutrino signals from PBHs as DM candidates, focusing on 14 dSphs observed by the IceCube Neutrino Observatory. Our analysis represents the most comprehensive investigation to date of memory-burdened PBH signals from these promising DM targets. By incorporating $\mathcal{D}$-factor measurements from four independent methodological approaches and employing a rigorous unbinned maximum-likelihood framework, we have derived robust constraints on the PBH abundance fraction across a broad mass range under some cases.

Our results demonstrate that the $k=1$ MB scenario yields significantly tighter constraints compared to the $k=2$ case, particularly for PBH masses between $10^7$ to $10^{10}$~g. This improvement stems from the favorable alignment between the harder neutrino spectra in the $k=1$ case and the optimal energy range of IceCube. Notably, for the $k=1$ case across all $\mathcal{D}$-factor sets, our analysis excludes PBHs as the dominant component of DM up to approximately $10^{10}$~g, substantially extending previous limits. For the $\mathcal{C}$ set, the constraints remain stronger than previous results up to $k \lesssim 1.5$, while for the other $\mathcal{D}$-factor sets, comparable improvements are achieved up to $k \lesssim 1.3$. The joint analysis of all 14 dSphs substantially enhances our sensitivity compared to individual source analyses, with improvements driven by the statistical leverage gained from combining independent measurements with uncorrelated systematics. While our constraints on lower mass PBHs remain less stringent than existing limits, our analysis shows increasingly competitive performance for higher PBH masses, highlighting the growing importance of neutrino-based searches for this parameter regime.

We provide a robust framework by combining multiple $\mathcal{D}$ factor datasets with a profile likelihood treatment of uncertainties, readily extendable to future dark matter searches with IceCube and next-generation neutrino telescopes. The upcoming IceCube-Gen2, with its significantly enhanced effective area and angular resolution, will greatly improve the sensitivity to neutrino signals from memory-burdened PBHs. This is particularly valuable for probing the $k=2$ scenario and extending the mass reach of our constraints. Our work thus establishes both the methodology and the foundation for future investigations, bringing us closer to understanding whether PBHs contribute to the dark matter content.

\section*{Acknowledgements}
Y.F.Z. is supported by the National Key R\&D Program of China (Grant No. 2017YFA0402204), the CAS Project for Young Scientists in Basic Research YSBR-006, and the NSFC (Grants No. 11821505, No. 11825506, and No. 12047503). J.Q.X. is supported by the NSFC (Grants No. 12473004).

\bibliographystyle{apsrev4-1}
\bibliography{references}

\end{document}